\documentclass[11pt]{article}
\usepackage{graphicx}
\usepackage{moriond,epsfig}
\newcommand{\BABARPubYear}    {01}

\newcommand{\BABARProcNumber} {27}
\newcommand{\SLACPubNumber} {8824}

\newcommand{\mybabar} { BaBar }
\input babarsym
\newcommand{\btopp}{\ensuremath{\Bz\to \pip\pim}}

\newcommand{\btoKpp}{\ensuremath{\Bz\to \Kp\pim}}

\newcommand{\btoKK}{\ensuremath{\Bz\to \Kp\Km}}

\newcommand{\btoppz}{\ensuremath{B^+\to \pip\piz}}

\newcommand{\btoKpz}{\ensuremath{B^+\to \Kp\piz}}

\newcommand{\btoKzK}{\ensuremath{B^+\to \Kzb\Kp}}

\newcommand{\btoKzpz}{\ensuremath{\Bz\to \Kz\piz}}

\newcommand{\etal}{{\it et al.}}

\def\fish    {\ensuremath{\cal F}}

\def\thsph    {\ensuremath{\theta_{\scriptscriptstyle S}}}

\long\def\inst#1{\par\nobreak\kern 4pt\nobreak
    {\it #1}\par\vskip 10pt plus 3pt minus 3pt}

\begin{document}
{\pagestyle{empty}

\begin{flushright}
SLAC-PUB-\SLACPubNumber \\
BABAR-PROC-\BABARPubYear/\BABARProcNumber \\
May, 2001 \\
\end{flushright}

\par\vskip 4cm

\begin{center}
\Large \bf MEASUREMENT OF CHARMLESS HADRONIC B DECAYS BRANCHING FRACTION AT BABAR.
\end{center}
\bigskip

\begin{center}
\large G.Cavoto \\ on behalf of the BaBar Collaboration  

Universit\'a di Roma La Sapienza, Dipartimento di Fisica and INFN,\\
  I-00185 Roma, Italy \\ E-mail: gianluca.cavoto@roma1.infn.it

\end{center}
\bigskip \bigskip

\begin{center}
\large \bf Abstract
\end{center}
We present preliminary measurements of the branching fractions  for charmless hadronic decays of $B$ mesons into two-body final states with  kaons, pions and a $\phi$ resonance.  The measurements are based on a data sample of approximately $23$ million $B \bar B $  pairs collected by the \mybabar detector at the \pep2\ asymmetric $B$ Factory at SLAC.

\vfill
\begin{center}
Contributed to the Proceedings of the \\
36th Rencontres De Moriond On QCD And Hadronic Interactions \\
17-24 Mar 2001, Les Arcs, France
\end{center}

\vspace{1.0cm}
\begin{center}
{\em Stanford Linear Accelerator Center, Stanford University, 
Stanford, CA 94309} \\ \vspace{0.1cm}\hrule\vspace{0.1cm}
Work supported in part by Department of Energy contract DE-AC03-76SF00515.
\end{center}

Charmless hadronic final states play
an important role in the study of \CP-violation.
In the Standard Model, all \CP-violating phenomena
are a consequence of a single complex phase in the Cabibbo-Kobayashi-Maskawa
(CKM) quark-mixing matrix \cite{ckm}.
  Measurements of the rates and \CP\ asymmetries for $B$ decays into the charmless final states $\pi\pi$ and $K\pi$ can be used to constrain the angles $\alpha$~\cite{gamalpha} 
of the unitarity triangle. Decays containing a $\phi$ meson are interesting since they are dominated by $ b \to s(d) \bar ss $ peguin diagram, with potential benefits  to estimates of  direct \CP-violation effects. They also allow an independent measurement of $\sin{2 \beta}$.

 We present  preliminary measurements of the branching fractions for
charmless hadronic decays of $B$ mesons in the final states $\pip\pim$,
$\Kp\pim$, $\Kp\piz$, $\Kz\pip$, $\Kz\piz$,  $\phi \Kp$, $\phi \Kz$, and $\phi K^{*+}(892)$ \cite{cc}.
The data sample used in these analyses  consists of $22.57\times 10^6$ \BB\ pairs,  collected at the \pep2\ \epem\ collider (SLAC) with the \mybabar detector \cite{babarnim} .

Hadronic events are selected based on track multiplicity and event topology.
 We use only good quality tracks: tracks are identified as pions or kaons using the Cherenkov angle $\theta_c$ measured by a unique, internally reflecting Cherenkov ring imaging detector (DIRC). Candidate $\KS$ mesons are reconstructed from pairs of oppositely charged tracks that
form a well-measured vertex and have an invariant mass within $3.5$ standard deviations ($\sigma$) of the nominal \KS\ mass \cite{PDG}. 
Candidate photons are defined as showers in the EMC that have the
expected lateral shape and are not matched to a track.
Candidate \piz\ mesons are formed from pairs of photons with an invariant mass within $3\sigma$ of the nominal \piz\ mass.  
The \piz\ candidates are then kinematically fitted with their mass 
constrained to the nominal \piz\ mass. 
All tracks (except \KS\ daughters) are required to have good-quality
$\theta_c$ measurements that are inconsistent with the expected value for a 
proton.  Electrons are rejected
based on specific ionization (\dedx) in the DCH system, shower shape
in the EMC,
and the ratio of shower energy to track momentum.  

Candidate $B$ mesons are reconstructed in the  various topologies: $h^+ h^{\prime -}$, $h^+\piz$, $\KS h^+$, $\KS\piz$,  $\phi h^+$, $\phi \KS$, and $\phi K^{*+}$ where the symbols $h$ and $h^{\prime}$ refer to $\pi$ or $K$. 
 For $\phi $ candidates, both tracks must be identified as kaons whose invariant mass must lie within a $\pm 30 \mevcc$ interval centered around $\phi$ mass. The selection of $ K^{*+}$ comprises $\Kp \piz$ and  $\KS \pip$ combinations within a $K \pi $ mass interval of $\pm 150 \mevcc$. A  $K^*$ helicity angle cut effectively requires $\piz$ momentum greater than $ 0.35  \gevc$.

Candidate B mesons are selected exploiting the  kinematic constraints provided by the \FourS\ initial state: we   define a energy-substituted mass $ \mes$, where $ \sqrt{s}/2 $ is  substituted for candidate's energy, and the difference $\Delta E$ between the B-candidate energy and  $ \sqrt{s}/2 $. 
For all modes the \mes\ resolution is dominated by the beam energy spread 
and is approximately $2.5\mevcc$, while $\Delta E $ resolution is mode dependent and dominated by momentum resolution.
 Candidates are selected in the range $5.2<\mes<5.3\gevcc$. Candidates
are accepted, depending on the decay topology, in  various $\Delta E$ ranges,  which  are restrictive enough to suppress backgrounds due to other 
types of $B$ decays.

The largest source of background is from random combinations of tracks and neutrals produced in the
$\epem\to \qqbar$ continuum (where $q=u$, $d$, $s$ or $c$).  In the CM
frame this background typically exhibits a two-jet structure. 
 In contrast, the low momentum and pseudoscalar nature of $B$ mesons in 
the decay \upsbb\ leads to a more spherically symmetric event.
  This topology difference is exploited using two event-shape quantities:
the angle $\thsph$~\cite{spheric} between the sphericity axes, evaluated in the CM frame, of the $B$ candidate 
and the remaining tracks and photons in the event.  The distribution of the 
absolute value of $ \cos\thsph $ is strongly peaked near $1$ 
for continuum events and is approximately uniform for \BB\ events.  We require
$|\cos\thsph| < 0.9$

The second quantity used in the analyses is a Fisher discriminant
${\cal F}$ which consists of a linear combination of 
several variables that distinguish signal from background~\cite{fisher}.
The experimental observables used
in the definition of \fish\ are the scalar sum of the momenta of all
tracks and photons (excluding the $B$ candidate daughters)
flowing into nine concentric cones centered on the thrust axis of
the $B$ candidate, in the CM frame.  Each cone subtends an angle of $10^\circ$
and is folded to combine the forward and backward intervals.

The global  detection efficiencies, which include the intermediate particle
branching fractions, are
listed in Table \ref{tab:brresults}.

\begin{table*}[!htb]
\begin{center}
\caption{
Summary of results for detection efficiencies ($\epsilon$), numbers of fitted signal yields ($N_S$), statistical significances, and measured branching fractions (\BR).  (  $90\%$ confidence level upper limits).
The efficiencies include the branching fractions for 
$\Kz\to\KS\to\pip\pim$, $\piz\to\gamma\gamma$, $\phi \to \Kp \Km$.  Equal branching fractions
for \upsbzbz\ and $\Bu\Bub$ are assumed.  
} 
\label{tab:brresults}
\begin{tabular}{lcccc} 
\hline\hline
Mode  & $\epsilon$ (\%) &  $N_S$ &  ~~~Stat. Sig. ($\sigma$)~~~ & 
\BR($10^{-6}$) \\ 
\hline
$\pip\pim$ &  $45$  & $41\pm 10$  & 
$4.7$  & $4.1\pm 1.0\pm 0.7$ \\
$\Kp\pim$ &  $45$  & $169\pm 17$  &
$15.8$ & $16.7\pm 1.6\pm 1.3$  \\
$\Kp \Km$  & $43$  & $8.2^{+7.8}_{-6.4}$   & $1.3$  & $<2.5$ \\
$\pip\piz$  & $32$  & $37\pm 14$  & $3.4$  & $<9.6$  \\
$\Kp\piz$    & $31$   & $75\pm 14$  & $8.0$  & $10.8^{+2.1}_{-1.9}\pm 1.0$  \\
$\Kz\pip$   & $14$  & $59^{+11}_{-10}$ & $9.8$  & $18.2^{+3.3}_{-3.0}\pm 1.7$  \\
$\Kzb\Kp$    & $14$  & $0.0^{+2.4}_{-0}$  & $0$    & $<2.5$  \\
$\Kz\piz$  & $9.6$   & $17.9^{+6.8}_{-5.8}$  & 
$4.5$  & $8.2^{+3.1}_{-2.7}\pm 1.1$  \\
$\phi \Kp$  & $18 $ & $31.4 ^{+6.7}_{-5.9}$ & $10.5$ & $7.7 ^{+1.6}_{-1.4}\pm 0.8$  \\
$\phi \pip$  & $19 $ & $0.9 ^{+2.1}_{-0.9}$ & $0.6$ & $\, < \, 1.4 $  \\
$\phi \Kz$  & $ 6$ & $10.8 ^{+4.1}_{-3.3}$ & $6.4$ & $8.1 ^{+3.1}_{-2.5}\pm 0.8$  \\
$\phi K^{*+}$  & $ 5 $ & $ $ & $4.5$ & $9.7 ^{+4.2}_{-3.4}\pm 1.7$  \\
\hline
$\phi K^{*+}_{K^+}$  & $ 2.5 $ & $ 7.1 ^{+4.3}_{-3.4} $ & $2.7$ & $12.8^{+7.7}_{-6.1}\pm 3.2$  \\
$\phi K^{*+}_{K^0}$  & $ 2.4 $ & $4.4 ^{+2.7}_{-2.0} $ & $3.6$ & $8.0 ^{+5.0}_{-3.7}\pm 1.3$  \\
\hline
\\
\hline\hline
\end{tabular}
\end{center}
\end{table*}
Signal yields are determined from an unbinned maximum likelihood fit using
\mes, $\Delta E$, \fish, $\theta_c$, $\phi$ mass, and $K^{*+}$ mass  (where applicable).
In each of the fits, 
the likelihood for a given candidate $j$ is obtained by summing the product of 
event yield $n_k$ and probability ${\cal P}_k$ over all possible signal and 
background hypotheses $k$.
The $n_k$ are determined by maximizing the extended likelihood function 
$\cal L$:
\begin{equation}
{\cal L}= \exp\left(-\sum_{i=1}^M n_i\right)\,
\prod_{j=1}^N \left[\sum_{k=1}^M n_k {\cal P}_k\left(\vec{x}_j;
\vec{\alpha}_k\right)
\right]\, .
\end{equation}
where ${\cal P}_k(\vec{x}_j;\vec{\alpha}_k)$ is the probability 
for candidate $j$ to belong to category $k$ 
(of $M$ total categories), based on its characterizing variables $\vec{x}_j$
and parameters $\vec{\alpha_k}$ that describe the expected distributions of these
variables.
The probabilities ${\cal P}_k(\vec{x}_j;\vec{\alpha}_k)$ are 
evaluated as the product of probability density functions (PDFs) for each of 
the independent variables $\vec{x}_j$, given the set of parameters 
$\vec{\alpha}_k$.
Monte Carlo simulated data is used to validate the
assumption that the fit variables are uncorrelated.
The exponential factor in $\cal L$ accounts for
Poisson fluctuations in the total number of observed events $N$.

The probabilities for each possible signal and background hypotheses are evaluated as the product of probability density functions (PDFs) for each of 
the independent variables.  The parameters of \mes, $\Delta E$, and \fish\ PDFs 
are determined from data, and are cross-checked with  Monte Carlo simulation.
 In particular, a sample of $D^*$-tagged $D^0 \rightarrow K^- \pi^+ $ decays is used to parametrize the $\theta_{C}$ distribution for pion and kaon tracks as a function of momentum. 
The results of the fits for the various  topologies are summarized in Table \ref{tab:brresults}.
We find evidence for the decay $B^+\to \pip\piz$ and 
measure a branching fraction of 
$\BR(\B^+\to\pip\piz) = (5.1^{+2.0}_{-1.8}\pm 0.8)\times 10^{-6}$.
However, the signal significance
is insufficient to claim observation. 
A $90\%$ confidence level upper limit is computed for this mode,
as well as for \btoKK\, \btoKzK , and $B^+ \rightarrow \phi \pip$ in the following manner. 
The upper limit on the signal yield for mode $k$ is given by the value of
$n_k^0$ for which
$\int_0^{n_k^0} {\cal L}_{\rm max}\,{\rm d}n_k/\int_0^\infty 
{\cal L}_{\rm max}\,{\rm d}n_k = 0.90$, where ${\cal L}_{\rm max}$ is 
the likelihood as a function of $n_k$,
maximized with respect to the remaining fit parameters.  

The result is then increased by the total systematic error.  The detection
efficiency is reduced by its systematic uncertainty in calculating the
branching fraction upper limit.  
 The statistical significance of a given
 channel is determined by fixing the yield to zero, repeating the fit,
and recording the change in $-2\ln{\cal L}$.

Figure~\ref{fig:prplots} shows the distributions in \mes\ 
 for events passing the selection criteria, as well as requirements on Fisher value, which are used to increase the relative 
 fraction of signal events
 of a given type. Fits to these distributions are overlaid.

\begin{figure}
\begin{center}
\psfig{figure=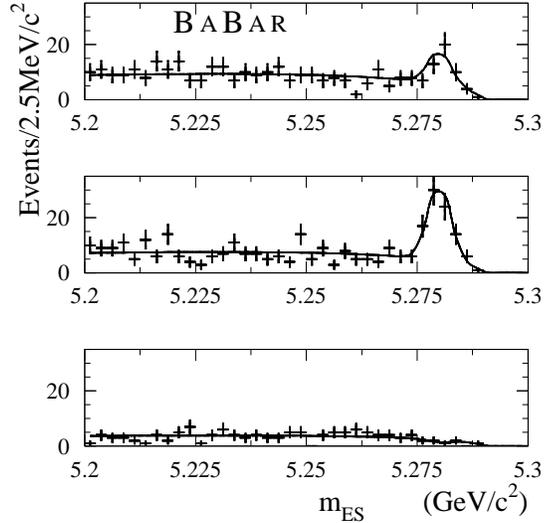,height=3in}
\caption{The \mes\ distributions for candidates which are (from top to bottom) 
$\pip\pim$-like, $\Kp\pim$-like,  $\Kp\Km$-like
The curves represent  fits to the distributions 
\label{fig:prplots}}
\end{center}
\end{figure}

Imperfect knowledge of the PDF shapes and of the detection efficiencies 
are the main sources of  systematic uncertainties on the  branching ratio
measurements.
Uncertainties in the PDF parameterizations
are estimated either by varying the PDF
parameters within $1\sigma$ of their measured uncertainties or by
substituting alternative PDF from independent control samples, and
recording the variations in the fit results.

In summary, we have measured branching fractions for the rare charmless
decays \btopp, \btoKpp, \btoKpz, $\Bu\to\Kz\pip$, \btoKzpz, $\phi \Kp$, $\phi \Kz$, and  $\phi K^{*+}$
and set upper limits on \btoKK, \btoppz, \btoKzK, and  $\phi \pip$.

\section*{Acknowledgments}
This work has been supported by the US Department of Energy and 
National Science Foundation, the Natural Sciences and Engineering
Research Council (Canada), the Institute of High Energy Physics (China), 
Commissariat
\`a l'Energie Atomique
and Institut National de Physique Nucl{\'e}aire et de Physique des Particules 
(France),
Bundesministerium
f{\"u}r Bildung and Forschung (Germany), Istituto Nazionale di Fisica 
Nucleare (Italy),
the Research
Council of Norway, the Ministry of Science and Technology of the Russian 
Federation and the
Particle Physics
and Astronomy Research Council (United Kingdom).


\begin{thebibliography}{99}
\bibitem{cc}
Charge conjugate states are assumed throughout, except where explicitly
noted.
\end{thebibliography}
\end{document}